\documentclass[conference]{IEEEtran}
\IEEEoverridecommandlockouts

\usepackage{amsmath,graphicx}
\usepackage{todonotes}
\usepackage{soul}

\usepackage{amssymb}
\usepackage{cite}
\usepackage{amsmath,amssymb,amsfonts}
\usepackage{algorithmic}
\usepackage{graphicx}
\usepackage{textcomp}
\usepackage{xcolor}
\def\BibTeX{{\rm B\kern-.05em{\sc i\kern-.025em b}\kern-.08em
    T\kern-.1667em\lower.7ex\hbox{E}\kern-.125emX}}

\begin{document}
\title{Acoustic identification of individual animals \\with hierarchical contrastive learning}


\author{
\IEEEauthorblockN{
Ines Nolasco\textsuperscript{$\dagger$} \quad 
Ilyass Moummad\textsuperscript{$\diamond$} \quad
Dan Stowell\textsuperscript{$\ddagger$} \quad
Emmanouil Benetos\textsuperscript{$\dagger$} 
}
\\
\IEEEauthorblockA{\textsuperscript{$\dagger$}\textit{Centre for Digital Music, Queen Mary University of London, London, UK}\\
\textsuperscript{$\diamond$}\textit{IMT Atlantique, CNRS, Lab-STICC, Brest, France}\\
\textsuperscript{$\ddagger$}\textit{Department of Cognitive Science and Artificial Intelligence, Tilburg University, Tilburg, Netherlands}\\
}

}

\maketitle

\begin{abstract}

Acoustic identification of individual animals (AIID) is closely related to audio-based species classification but requires a finer level of detail to distinguish between individual animals within the same species. In this work, we frame AIID as a hierarchical multi-label classification task and propose the use of hierarchy-aware loss functions to learn robust representations of individual identities that maintain the hierarchical relationships among species and taxa.
Our results demonstrate that hierarchical embeddings not only enhance identification accuracy at the individual level but also at higher taxonomic levels, effectively preserving the hierarchical structure in the learned representations. By comparing our approach with non-hierarchical models, we highlight the advantage of enforcing this structure in the embedding space.
Additionally, we extend the evaluation to the classification of novel individual classes, demonstrating the potential of our method in open-set classification scenarios.
\end{abstract}
\begin{IEEEkeywords}
Bioacoustics, contrastive loss, hierarchical classification, representation learning, open set.
\end{IEEEkeywords}
%


\section{Introduction}
\label{sec:intro}

Acoustic identification of individual animals (AIID) refers to the automatic differentiation of vocalisations among different individuals within a group. Many animal species exhibit distinct Acoustic Signatures —unique vocal characteristics that can be leveraged for identification. Traditionally, AIID has been approached in scenarios limited to a single species or a small, predefined group of individuals, \cite{Sadhukhan2021, Yin2004, wijers2021vocal}. However, this approach restricts the applicability of AIID in real-world settings where multiple species coexist and the set of individuals is not fixed, \cite{Ptacek2016, ntalampiras2021acoustic, knight2024individual}.
The AIID problem across multiple species and taxa can be framed as a constrained multi-label classification task, where each instance requires the prediction of three labels: individual identity, species, and taxon. Importantly, there is a constraint such that knowledge of a label at a lower level (e.g., a specific individual) inherently provides information about the higher levels.

We propose that hierarchical classification is particularly well-suited for animal-related acoustic tasks due to the strong phylogenetic relationships that influence their physical and behavioural traits, \cite{arato2021phylogenetic}. In this context, several related bioacoustic tasks can be organised within a hierarchical structure, akin to animal taxonomy. For example, while species classification and AIID both seek to distinguish vocal characteristics, they differ primarily in the granularity of detail required. Species classification focuses on identifying differences between species, whereas AIID targets the distinction of individuals within a species. This concept extends to broader taxonomic classifications, such as differentiating between mammals and birds, which generally requires more generalised characteristics. Nevertheless, all these tasks share a common foundation in signal representation, with the critical difference being the level of detail at which the algorithm analyses the signals.
Hierarchical classification is often implemented by constraining the possible classes at lower levels of the hierarchy based on the prediction of the previous level  \cite{taxonet, chang2021your}. 
Here instead, the hierarchical structure in the label space is leveraged to guide the learning of the embeddings and highlight the similarities and the hierarchical relationships between features. By utilising hierarchical information at the feature level, we are able to obtain a representation of each ID that preserves the information regarding species, and taxonomic group. Preserving the ``full picture" of our data examples is an important step towards classification of previously unseen classes.

Distance-based methods are particularly suitable in open-set scenarios, as they enable the system to learn data representations that can accommodate the inclusion of novel classes. To this end, we adapt the hierarchical contrastive learning approach proposed in \cite{hiersupcon} to AIID, aiming to create a more generalisable AIID system. In this implementation, the multi-label constraint is enforced by producing one prediction for each level of the hierarchy considered. Further refinements address the confidence level at each hierarchical level, given the confidence in the lower levels.

The primary contributions of this work are: 1) To the best of our knowledge, this is the first work to apply contrastive learning to AIID, leveraging hierarchical contrastive loss to create robust representations for this task.
2) Evaluation of hierarchical and contrastively-trained embeddings for the open-set scenario in which new ``leaf" classes are encountered. 

\section{Related Work}
\label{sec:rwork}

Central to AIID in natural contexts are the challenges of generalisation and open set classification. Recently, this task has gained attention due to advancements in deep learning and foundation models for bioacoustics \cite{knight2024individual, hagiwara2023aves}. Leveraging the hierarchical structure of labels to improve AIID, was first explored in  \cite{nolasco2022rank}, who proposed a hierarchy-aware loss function to guide the learning of embeddings that preserve the hierarchical relationships and enhance individual classification across species. Hierarchy can also be useful in the open set classification, as shown in \cite{lang2024coarse}.

In this work, we adopt the hierarchical contrastive learning loss from \cite{hiersupcon} to further these goals.
Contrastive learning has emerged as a powerful approach for learning robust feature representations. In its self-supervised form~\cite{simclr}, contrastive learning aims to minimise the distance between positive pairs of samples, typically augmentations of the same instance, while maximising the separation from negative pairs, enabling models to learn meaningful feature spaces without explicit labels. This paradigm has proven effective in a variety of domains~\cite{cookbookssl}. The extension to supervised contrastive learning~\cite{supcon} incorporates label information to further enhance discriminative power by grouping samples from the same class closer together, thus improving the learned feature space for tasks such as classification. Supervised contrastive learning has shown promising results in bioacoustics for few-shot capabilities in classification~\cite{sslbirds} and detection~\cite{dcasemoummad}.

Recently, hierarchical contrastive learning~\cite{hiersupcon} has emerged as an advancement in visual representation learning, addressing the need for models to recognise not only coarse-level categories but also fine-grained hierarchical relationships within the data. By incorporating hierarchical structures, such methods can create more structured and context-aware feature spaces, which are particularly useful for tasks where data exhibits multi-level or nested class relationships, as often seen in bioacoustics.

\vspace{-0.1cm}
\section{Dataset}
\label{sec:data}
The dataset is a collection of short recordings from vocalisations of animals of different species (see summary in table \ref{tab:dataset}). 
The data is sourced from different research initiatives focusing on animal communication. Recordings are made in the animal's natural environment,  by experts that follow the individuals and annotate their ID. Due to the use of different acquisition methods, the recordings present high variation in acoustic characteristics.  
The dataset represents a natural setting in which systems for AIID need to operate. 


Data is split into training, validation and test sets, which contain examples of 66 individuals from the various species. Additionally, an unseen ID set is defined which contains 3 novel IDs for each of the species in the training set (21 novel ID classes).  In total there are 6055 examples for training, 1529 for validation, 1912 for test and 4799 examples of unseen ID classes.

\begin{table}[h!]
    \centering
    \begin{tabular}{llllll}
    \textbf{Species} & \textbf{Taxon} & \# \textbf{Ids}  \\
    \hline
    Chiffchaffs (CHF)\cite{stowell_2018_1413495} & Birds  & 23 \\
    Tree pipits (TP)\cite{stowell_2018_1413495} & Birds  & 10 \\
    Little Owls (LO)\cite{stowell_2018_1413495} & Birds  & 16 \\
    Eurasia eagle owls (EEO) & Birds& 7 \\
    Spotted hyenas (SH)\cite{lehmann2022long} & Mammals  & 5 \\
    Hyrax (HY)\cite{demartsev2019lifetime}& Mammals &  19  \\
    Grey wolves (GW)\cite{root2013improving} & Mammals &  7\\
    \hline \hline
\textbf{Total number of recordings }& 14295 & \\ \hline
    \end{tabular}
    \caption{Summary of dataset.  }
    \label{tab:dataset}
\end{table}


\section{Methods}
\label{sec:method}


\subsection{Representation learning}

In contrastive learning approaches, the model architecture comprises two important components: A feature extractor designed to map the input data into an abstract latent representation; and a shallow neural network called projector, which projects the features to a low dimensional space where the contrastive loss is computed. The projector is primarily used to train the feature extractor and is discarded after training (Fig.~\ref{fig:framework}). Here, we describe the supervised contrastive losses explored in our experiments.

\subsubsection{Supervised contrastive loss}

The supervised contrastive loss (SupCon)~\cite{supcon} is formulated as:
\begin{equation}
    L_{\text{SupCon}} = \sum_{i \in I} \frac{-1}{|P(i)|} \sum_{p \in P(i)} \log \frac{\exp(\boldsymbol{z}_i \cdot \boldsymbol{z}_p / \tau)}{\sum_{n \in N(i)} \exp(\boldsymbol{z}_i \cdot \boldsymbol{z}_n / \tau)}
\end{equation}
where \( i \in I \) indexes an augmented sample within a batch, \( P(i)={\{p\in I:{{y}}_p={{y}}_i}\} \) is the set of positive samples sharing the same label as \( i \), \( N(i) = I\setminus \{i\} \) represents the set of negative samples, and \( \tau \) is a temperature scaling parameter.

\subsubsection{Hierarchical Multi-label Contrastive Learning}

SupCon~\cite{supcon} only leverages one hierarchy level. To leverage hierarchical class structures, we adopt a hierarchical multi-label contrastive learning framework that incorporates hierarchy-aware losses~\cite{hiersupcon}: 

\textbf{HiMulCon} (Hierarchical multi-label contrastive loss):
\begin{equation}
\label{eq:himulcon}
    L_{\text{HiMulCon}} = \sum_{l \in L} \frac{1}{|L|} \sum_{i \in I} \frac{-\lambda_l}{|P_l(i)|} \sum_{p \in P_l(i)} L_{\text{pair}}(i, p_i^l)
\end{equation}
where \( L \) represents different levels in the hierarchy, \( \lambda_l \) is a level-dependent penalty factor, \( P_l(i) \) is the set of positive pairs at level \( l \), and \( L_{\text{pair}} \) calculates the contrastive loss for a specific pair.

\subsubsection{Hierarchical Multi-label Contrastive constraint enforcing}
An additional constraint is included into the previous loss in order to 
ensure that losses at each level do not decrease with increasing hierarchy depth:

\begin{equation}
    L_{\text{HiMulConE}} = \sum_{l \in L} \frac{1}{|L|} \sum_{i \in I} \frac{-\lambda_l}{|P(i)|} \sum_{p \in P_l(i)} \max(L_{\text{pair}}(i, p_i^l), L^{\text{pair}}_{\text{max}}(l-1))
\end{equation}
the term  \( L^{\text{pair}}_{\text{max}} (l-1) \) is the maximum loss computed from the previous level and defines the absolute minimum value of loss each level can achieve.
HiMulCon consists on an independent penalty defined on each level, whereas the added constraint is a dependent penalty that is defined in relation to the losses computed at the lower levels. Both combined form the \textbf{HiMulConE} loss.

\subsection{Classification} 
\label{sec:classif}

We assess the effectiveness of the feature extractors learned using various loss functions with a $k$-Nearest Neighbor (kNN) classifier, a common evaluation protocol for learned representations~\cite{cookbookssl}. The evaluation is conducted in two scenarios: 1) classification of new examples from the same classes used during training, and 2) classification of data from previously unseen classes.


In the first scenario, for each test sample, the predicted label is determined by finding the $k$ nearest neighbors from the training set and assigning the majority label from among those neighbors. This method allows us to evaluate how well the learned features generalize to new samples within the same classes.

In the second scenario, we assess the ability of the learned representations to generalize to previously unseen classes. Here, we consider both the training examples and all examples from the unseen classes when determining the nearest neighbors for a test sample. This approach allows us to evaluate how new classes are represented in the learned feature space, even though the model has not encountered them during pre-training.
To further assess the generalisation capabilities of the models in an open-set scenario, we implement a one-shot classification setting. In this setup, a support set containing only one example from the unseen classes is used to classify query samples. The nearest neighbors for each query are identified from both the support set and the training set (with the latter acting as distractors). This evaluation highlights the ability of the models to learn from very few labeled examples of new classes in a more realistic scenario.

\section{Experiments}
\label{sec:experiments}
\subsection{Experimental setup}
\begin{figure}[h!] 
    \centering
    \includegraphics[width=0.5\textwidth]{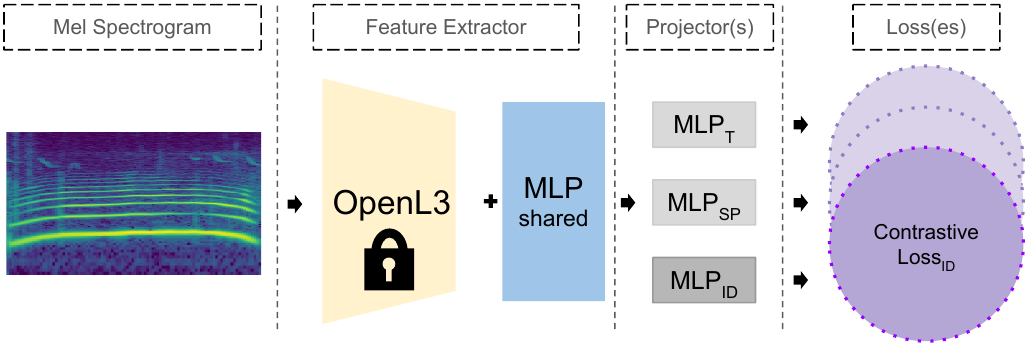} 
    \caption{Overview of our pretraining pipeline: audio recordings of animal calls are first processed through a frozen, pretrained OpenL3 model to extract high-level representations on each 25ms segment of the audio file. the final embeddings is the average of these across the whole call. The final openl3 embedding is then passed through a MLP to adapt the features specifically for bioacoustic sounds. The adapted features are subsequently fed into a projector to perform supervised contrastive learning for individual identification (ID). For hierarchical contrastive learning, two additional projectors are included for the species (SP) and taxa (T) classification.}
    \label{fig:framework}
\end{figure}

The losses are applied into the training of the network described in Fig. \ref{fig:framework}. This is defined as a hierarchical network, which contains one head for each level of the hierarchy.
Embeddings from the OpenL3 pretrained model are computed using available code\footnote{https://github.com/hearbenchmark/hear-baseline} from the HEAR challenge \cite{turian2022hear}.


The experiments are defined as: 
\begin{enumerate}
    \item \textbf{SC:} for the non-hierarchical baseline we train a modified version of the network in figure\ref{fig:framework}, where only the ID projector is kept. This network is trained with the SupCon loss, as defined in eq.(1).

    \item \textbf{HC:} Hierarchy is included by training the complete network with the HiMulCon loss function (eq.(2)). 
    In this experiment we define equal contribution of each level of the hierarchy into the final loss. ($\lambda_l$ =1/3)

    \item \textbf{HC$\lambda$:} In order to test the effect of the combining factor $\lambda _l$ of eq.(2) on the performance, we experiment with various values. 
    Fine-tuning on validation set shown the best values as $\lambda _{ID} = 10$, $\lambda _{species} = 5$, and $\lambda _{taxa} = 1$.

    \item \textbf{HCE:} the hierarchy is further enforced on the learning process by applying the HiMulConE loss function (eq.(4)). 

    \item \textbf{HCE$\lambda$:} Similar to experiment 3) various values for $\lambda _l$ are experimented with and through tuning on the validation set we select the combination $\lambda _{ID} = 10$, $\lambda _{species} = 1$, and $\lambda _{taxa} = 1$.
\end{enumerate}

Several hyperparameters, such as temperature, learning rate, weight decay, batch size, and $\lambda$ values, are tuned using a sweep process across a defined range of values. The best parameters are selected which produce the highest accuracy on the validation set. For validation, accuracy is measured by applying \textit{KNN} with $k=1$ nearest neighbours to the embeddings extracted from the shared layer of the network. 

\subsection{Evaluation}

The evaluation process is designed to assess three key aspects of the models: 1) How effectively can the models classify individual IDs; 2) Do the learned embedding spaces correctly capture the hierarchical relationships between labels? And 3), Can the models generalise to unseen classes at the ID level?
To these effects, first the trained models are evaluated for their classification accuracy across all three levels of the hierarchy. Classification is performed by applying the \textit{KNN} classification (see section \ref{sec:nnclass}) to the embeddings extracted from the shared layer of the network (see Fig. \ref{fig:framework}).
Due to the imbalanced nature of the dataset, the accuracy values are computed using the balanced\_accuracy\_score implementation from \textit{sklearn}~\cite{brodersen2010balanced}. This ensures that classes with fewer examples do not disproportionately affect the overall performance evaluation. 

Secondly, We define two types of hierarchical inconsistency errors: \textbf{ species/ID} - predicted ID does not belong to the predicted species; \textbf{taxon/species} - predicted species does not belong to the predicted taxonomic group. 
Here the evaluation focus not on how correct the model is, but instead if it is consistent in its predictions accordingly to the hierarchy.      

And finally we assess the generalisation capability to classify unseen ID classes in two ways, first we evaluate if the trained embedding space is suitable to represent novel classes by employing NN classification using the combination of train and unseen ID set as the reference set, (see section \ref{sec:classif}). Secondly, because this is not a real use case of classification on novel classes, we also test the models' ability to identify a new class based on one single example. The 1-shot classification results allow us to understand the potential use of these models in the open set scenario.

\subsection{Results}
\label{sec:results}
All the trained models are evaluated on both test and unseen ID set. Accuracy values are reported in Tables \ref{tab:res_test} and \ref{tab:res_useenid} respectively. 
Regarding the analysis on hierarchical inconsistency errors, results obtained indicate that all models produce consistent predictions regarding the hierarchy. Since no errors of these types were found we are led to observe that all misclassifications occur within the correct parent class for the ID and Species levels.

\begin{table}[h!]
\resizebox{\columnwidth}{!}{
\centering
\begin{tabular}{rcccccc}
\textbf{Test Set}  & \textbf{SC} & \textbf{HC} & \textbf{HC$\lambda$} & \textbf{HCE} & \textbf{HCE$\lambda$} \\ \hline
\textbf{CHF} & & & & &  \\ 
\quad Species & 99.8 & \textbf{100} & \textbf{100} & \textbf{100} & \textbf{100} \\
\quad ID & 90.9 & 94.4 & \textbf{94.7}& 94.4 &93.8  \\ \hline
\textbf{TP} & & & & &  \\
\quad Species & 92.8 & \textbf{100} & 98.7 & 97.8 & 98.9 \\
\quad ID &  56.8 & \textbf{73.7} & 62.8 & 67.8 & 69.9 \\\hline
\textbf{LO} & & & & & \\
\quad Species &  98.6 & \textbf{100} & \textbf{100} & 99.3 & \textbf{100} \\
\quad ID & 44.7 & \textbf{70.3} & 69.7& 53.3 & 69.7 \\\hline
\textbf{EEO} & & & & & \\
\quad Species & 98.1  & 98.0 & 98.1 & 98.1 & \textbf{100}  \\
\quad ID & 53.8 & \textbf{55.8} & 61.5 & 44.2 &53.8 \\ \hline
\textbf{SH} & & & & &  \\
\quad Species & 98.8 & \textbf{100} & \textbf{100} & \textbf{100} & 96.6  \\
\quad ID & 96.5 & \textbf{100} & 97.7 & 98.9 & 94.3  \\\hline
\textbf{HY} & & & & &  \\
\quad Species & 99.4 & \textbf{100} & \textbf{100} & \textbf{100} & 99.4 \\
\quad ID & 49.7 & 57.1 & \textbf{60.4} & 44.6 & 59.9\\\hline
\textbf{GW} & & & & &  \\
\quad Species & \textbf{100} & \textbf{100} & \textbf{100} & \textbf{100} & \textbf{100} \\
\quad ID & 92.3 & 88.4 & \textbf{96.1} &  92.7 & 92.3 \\\hline
\textbf{Balanced acc} & & & & & \\
\quad Taxon & 99.7 & \textbf{100}& \textbf{100} & \textbf{100} &  99.8 \\
\quad Species & 98.2 & \textbf{99.7} & 94.5 & 99.3 & 99.3 \\
\quad ID & 61.0 & \textbf{73.2} & 72.2 & 64.2 & 72.3 \\ \hline
\end{tabular}
}
\vspace{+0.1cm}
\caption{Balanced accuracy, overall and for each species, on the test set across all evaluated models. }
\label{tab:res_test}
\end{table}


\begin{table}[h!]
\resizebox{\columnwidth}{!}{%
\centering
\begin{tabular}{lcccccc}
\textbf{Unseen IDs} & \textbf{SC} & \textbf{HC} & \textbf{HC$\lambda$} & \textbf{HCE} & \textbf{HCE$\lambda$} \\ \hline
\textbf{NN} & & & & &  \\
\quad Taxon   & 99.1 &  \textbf{99.7} & \textbf{99.7} & 99.6 & 99.1 \\
\quad Species & 96.9 & \textbf{99.1} & \textbf{99.1 }& 98.3  & 97.8  \\
\quad ID      & 80.9 & 85.7          & \textbf{88.1} & 86.0 & 84.7\\
\hline
\textbf{1-shot} & & & & &  \\
\quad Taxon & 92.2 & \textbf{97.7} & 96.5 & 97.3 & 95.6 \\
\quad Species & 84.2 & 92.3 & \textbf{93.8 }& 89.3 & 88.2 \\
\quad ID & 6.0 & 9.6 & \textbf{15.3}& 9.8 & 13.22 \\ 
\hline
\end{tabular}
}
\vspace{+0.1cm}
\caption{Balanced accuracy results on the unseen ID set following two classification processes: NN classification and 1-shot classification.}
\label{tab:res_useenid}
\end{table}



\vspace{-0.5cm}
\section{Discussion}
In this work, we framed the problem of acoustic identification of individuals as a multi-label hierarchical task. Through the use of contrastive learning-based methods, we investigated how learning an embedding space that captures the hierarchical structure of the labels can enhance individual identification. Additionally, we explored the limits of these models as encoders for performing classification of novel classes at the ID level.

The accuracy results on the test set (see table \ref{tab:res_test}) demonstrate that guiding the models to jointly optimise distances between embeddings across all hierarchical levels improves performance, not only at the fine-grained ID level but also at higher levels of the hierarchy, such as taxon and species. This confirms the value of preserving hierarchical relationships in the learned representations. Additionally, per-species accuracy results indicate the heterogeneous nature of the problem which contributes to the challenge of performing AIID for multiple species. 

When the models are applied to novel ID classes, the results indicate that the new classes are well represented in the learnt embedding space. Also the advantage of applying hierarchy aware training losses is clear in this novel class scenario. 
In a real scenario, a novel ID class scenario presents more closely as an open set classification problem, where we need to identify a new class from the first moment an example appears without having a complete representation from other examples of the new class. This situation is approximated here by employing a 1-shot classification approach. All the models exhibited a notable drop in accuracy, however showing a clear advantage of the models that included hierarchy. Furthermore, despite this decline, the models consistently preserved the hierarchical structure, as evidenced by the strong performance at the taxon and species levels. Moreover, the absence of consistency errors suggests that most misclassifications occur within the correct parent class—errors at the ID level involve confusion between IDs of the same species, rather than across species.

Overall, this work shows the potential of hierarchical embeddings for improving identification accuracy across multiple levels of an hierarchical problem. And while few-shot learning remains a challenging task, especially for novel classes at the individual ID level, the preservation of hierarchical integrity indicates that our approach provides a robust framework for AIID. 
Although we employ frozen features in this work, future research could investigate feature adaptation techniques
to enhance classification performance and explore methodologies for open-set classification.

\vspace{-0.2cm}
\section*{Acknowledgements} \label{sec:acknowledgement}
 I. Nolasco is supported by the Engineering and Physical Sciences Research Council [grant number EP/R513106/1]. E. Benetos is supported by a RAEng/Leverhulme Trust Research Fellowship [grant number LTRF2223-19-106]. 




\bibliographystyle{IEEEbib}
\bibliography{strings,refs}

\end{document}